\documentclass[manuscript]{acmart}
\AtBeginDocument{%
  \providecommand\BibTeX{{%
    \normalfont B\kern-0.5em{\scshape i\kern-0.25em b}\kern-0.8em\TeX}}}

\usepackage{subfigure}
\usepackage{subcaption}
\usepackage{graphicx}

\setcopyright{acmlicensed}
\copyrightyear{2024}
\acmYear{2024}
\acmDOI{XXXXXXX.XXXXXXX}

\acmConference[CHI '24]{ACM conference on Human Factors in Computing Systems}{May 11--16, 2024}{Hawaii, USA}

\acmBooktitle{CHI '24: ACM conference on Human Factors in Computing Systems,
 May 11--16, 2024, Hawaii, USA} 
\acmISBN{978-1-4503-XXXX-X/18/06}

\begin{document}

\title{User-Driven Adaptation: Tailoring Autonomous Driving Systems with Dynamic Preferences}

\author{Mingyue Zhang}
\affiliation{%
  \institution{Southwest University}
  \city{Chongqing}
  \country{China}}
\email{myzhangswu@swu.edu.cn}

\author{Jialong Li}
\affiliation{%
  \institution{Waseda University}
  \city{Tokyo}
  \country{Japan}}
\email{lijialong@fuji.waseda.jp}

\author{Nianyu Li}
\authornote{Corresponding author}
\affiliation{%
  \institution{ZGC Laboratory}
  \city{Beijing}
  \country{China}}
\email{li\_nianyu@pku.edu.cn}

\author{Eunsuk Kang}
\affiliation{%
  \institution{Carnegie Mellon University}
  \city{Pittsburgh}
  \country{USA}}

\author{Kenji Tei}
\affiliation{%
  \institution{Tokyo Institute of Technology}
  \city{Tokyo}
  \country{Japan}}
\email{tei@c.titech.ac.jp}

\renewcommand{\shortauthors}{Zhang and Li, et al.}

\begin{abstract}

In the realm of autonomous vehicles, dynamic user preferences are critical yet challenging to accommodate. Existing methods often misrepresent these preferences, either by overlooking their dynamism or overburdening users as humans often find it challenging to express their objectives mathematically. The previously introduced framework, which interprets dynamic preferences as inherent uncertainty and includes a ``human-on-the-loop'' mechanism enabling users to give feedback when dissatisfied with system behaviors, addresses this gap. In this study, we further enhance the approach with a user study of 20 participants, focusing on aligning system behavior with user expectations through feedback-driven adaptation. The findings affirm the approach's ability to effectively merge algorithm-driven adjustments with user complaints, leading to improved participants' subjective satisfaction in autonomous systems.

\end{abstract}
\keywords{Autonomous Driving, Preference Adaptation, Human on the Loop}

\maketitle

\section{Introduction}
In the evolution of software systems, particularly in areas like autonomous vehicles and cyber-physical systems, the need to understand and prioritize user preferences has become paramount. User preferences typically involve determining the relative importance of conflicting quality attributes, such as efficiency, safety, and privacy~\cite{nianyuPreference}. However, one-size-fits-all or standardized approaches often don't adequately meet individual user needs, leading to dissatisfaction and anxiety~\cite{whatdriverwants,passengercomfortAnxiety}. Thus, integrating user preferences effectively is crucial in user-centric design.

User preferences are dynamic and context-dependent. 
Consider an autonomous vehicle: initially, a user might prioritize efficiency, favoring the shortest routes to their destination. However, in certain scenarios, such as during a vacation, their emphasis might move towards ride comfort, favoring smoother and more leisurely routes to enjoy the view, even if they are longer. 
Similarly, customizable self-tracking tools permit frequent adjustments to display aesthetics and functionalities~\cite{preferencePersonalTracking} catering to unique user needs, with some users making changes multiple times daily. This variability underlines the necessity for systems to adapt their decision-making strategies swiftly to align with evolving user expectations~\cite{DBLP:conf/seams/WohlrabMV22,percomworkshoppreference,9391909}.

Existing research in autonomous driving often misrepresent user preferences, either by pre-setting preferences before deployment without considering individual variability and dynamism \cite{DBLP:conf/chi/DillenILNCS20}, or by assuming that users can clearly understand and articulate their preferences \cite{seamskenote2021, articleneuroscience, DBLP:conf/dagstuhl/2009adaptive}. Addressing these issues, \citet{nianyuPreference} developed a novel 'human-on-the-loop' framework to accommodate the inherent uncertainty in dynamic human preferences. This framework was inspired by the recognition that users are adept at identifying and expressing dissatisfaction with system behaviors \cite{DBLP:conf/hotos/SoulesG03, seamskenote2021}. 
Central to this framework is the incorporation of real-time user dissatisfaction, expressed as complaints, into the fitness function of a genetic algorithm. This algorithm is designed to detect and adapt to nuanced changes in user preferences over time. Consequently, the system's decision-making process is continuously updated and refined, by an evolving understanding of user preferences, thereby enhancing the alignment between system behaviors and user desires. However, its effectiveness of this work in adapting to changing human preferences and optimizing system performance has only been validated through theoretical experiments. These experiments, while informative, were conducted with predefined preferences and did not involve direct user interaction, which presents a limitation in terms of validating the framework's practical applicability and effectiveness in real-world scenarios.

Expanding upon the established theoretical foundation, this work applies the previously discussed framework to an autonomous driving system, incorporating a practical user study with 20 participants. This study aims to address a critical gap - the real-world application and validation. 
This study concentrates on essential quality attributes, including efficiency, riding comfort, and landscape aesthetics. Utilizing the Unreal Engine Simulator~\cite{UnrealEngine2024}, we created several 3D autonomous driving scenarios to evaluate the framework's effectiveness in harmonizing system operations with user expectation, particularly in the context of refining route selections based on participant feedback. The results of our user study clearly illustrate that the framework effectively aligns algorithm-derived system behaviors with self-reported user preferences, while also significantly enhancing user satisfaction and reducing the frequency of complaints about system behaviors. This improvement is a testament to the framework's ability to dynamically adjust to user preferences in real-time scenarios, offering a more personalized and satisfying user experience in autonomous driving.


The rest of the paper is organized as follows:
Section~\ref{backgroundrelWork} delves into related work.
Section~\ref{sec:scenario} lays out an exploration scenario centered on route choice in autonomous driving scenario, and provides a brief introduction of the preference adaptation approach in use. 
Section~\ref{sec:userstudy} details our practical user study, followed by the study results and an in-depth analysis in Section~\ref{sec:results}.
Section~\ref{sec:conclusion} concludes the paper, touching on limitations and possible avenues for future exploration.


\section{Related Work on Autonomous Driving and Preference Adaptation}
\label{backgroundrelWork}

The field of autonomous driving encompasses various challenges and developments. Chu et al. examine the role of safety drivers in the autonomous vehicle industry, emphasizing their impact on risk management and professional development~\cite{DBLP:conf/chi/ChuZSGLGDZ23}. 
Tener and Liu highlight the ongoing need for human assistance in autonomous vehicles (AVs), despite technological advancements~\cite{DBLP:conf/chi/TenerL22}. Schneider et al. explore how system transparency in AVs affects user experience and safety perception, providing design guidelines for integrating user experience with autonomous driving~\cite{DBLP:conf/chi/SchneiderHRGTG21}.  Dillen et al. focus on how fixed driving styles in AVs can conflict with passenger expectations, affecting comfort and anxiety~\cite{DBLP:conf/chi/DillenILNCS20}. These studies collectively highlight the diverse aspects of autonomous driving, from user experience to system design.

In the rapidly evolving landscape of transportation intelligence, the necessity to understand and integrate driver preferences into decision-making scenarios for AVs is becoming increasingly critical. 
 Pan et al. utilized inverse reinforcement learning (IRL) to understand taxi driver preferences in their passenger-search actions, offering insights into the dynamic nature of driver preferences~\cite{driverPreference}.  In typical autonomous system development, users are required to specify their preferences by ranking various quality attributes before the system's deployment~\cite{DBLP:conf/models/SongBCC13,DBLP:conf/seams/WohlrabMV22}. However, a significant challenge arises as users often struggle to precisely quantify these preferences in numerical terms~\cite{seamskenote2021,articleneuroscience,DBLP:conf/dagstuhl/2009adaptive,9723811,Abe2024Enhancing}, a critical aspect in customizing AV algorithms for individual user experience.


To emphasize the dynamic adaptation of preferences during runtime, adjusting utility functions to mirror the preference evolution is key~\cite{percomworkshoppreference}. Interactive techniques that assist users in adjusting rankings and understanding decision impacts enhance this adaptability~\cite{DBLP:journals/tvcg/GratzlLGPS13,DBLP:conf/cikm/KuhlmanVDNDPRH18,DBLP:journals/tvcg/PajerSTSMP17}. 
Wohlrab et al. and Song et al. have each contributed to this area by focusing on weight adjustments in response to contextual changes or direct user inputs~\cite{DBLP:conf/seams/WohlrabMV22,DBLP:conf/models/SongBCC13}. A notable study in AVs proposed voice-guided input for drivers to express their driving style preferences, highlighting the importance of integrating user preference in AV control systems~\cite{kim2021guiding}. These studies operate under the assumption that users have a clear understanding of their preferences and can articulate them in mathematical terms.

\section{Exploration Scenario and GA-based Preference Adaptation}
\label{sec:scenario}

\begin{figure}[h!tbp]
    \vspace{-0.2cm}
    \centering
    \subfigure[Building map for motivating scenario.]{
        \includegraphics[width=0.67\textwidth]{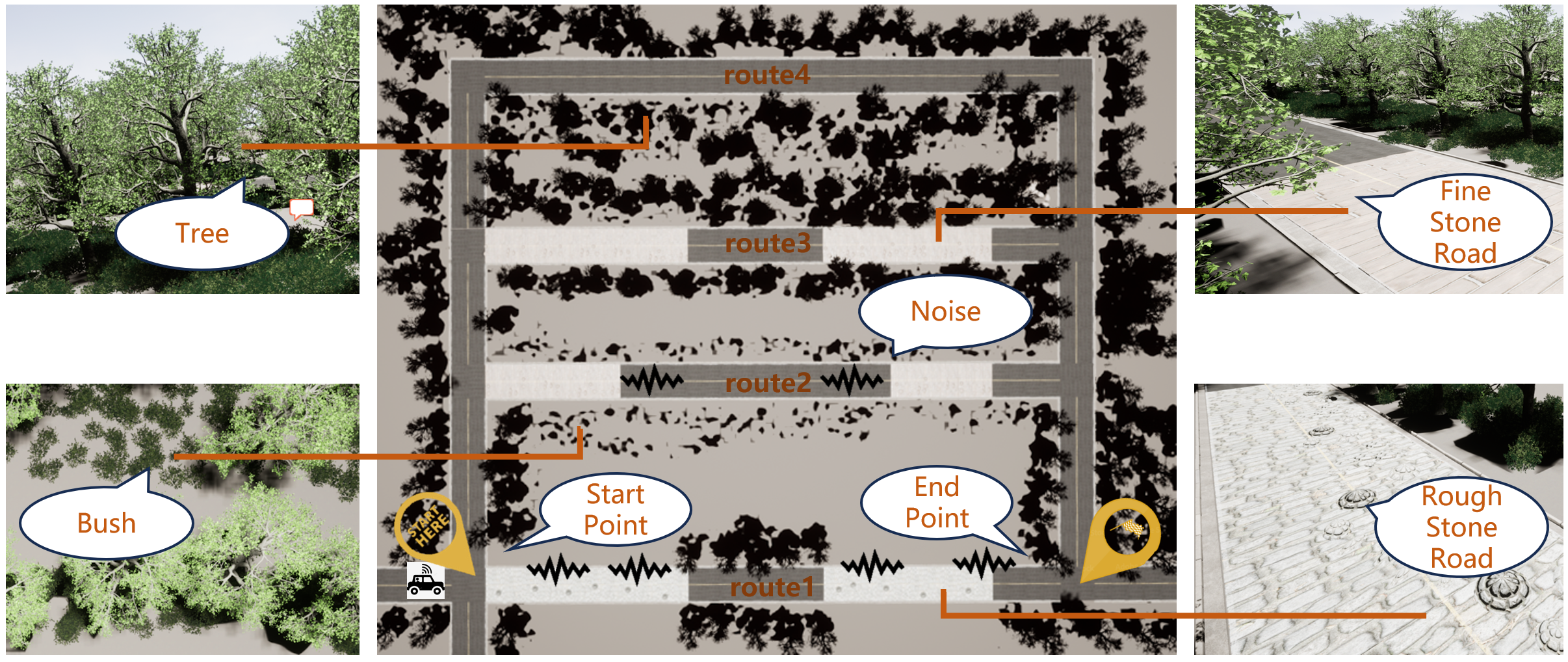}
    }
    \quad 
    \subfigure[Screenshots of auto-driving video.]{
    \includegraphics[width=0.28\textwidth]{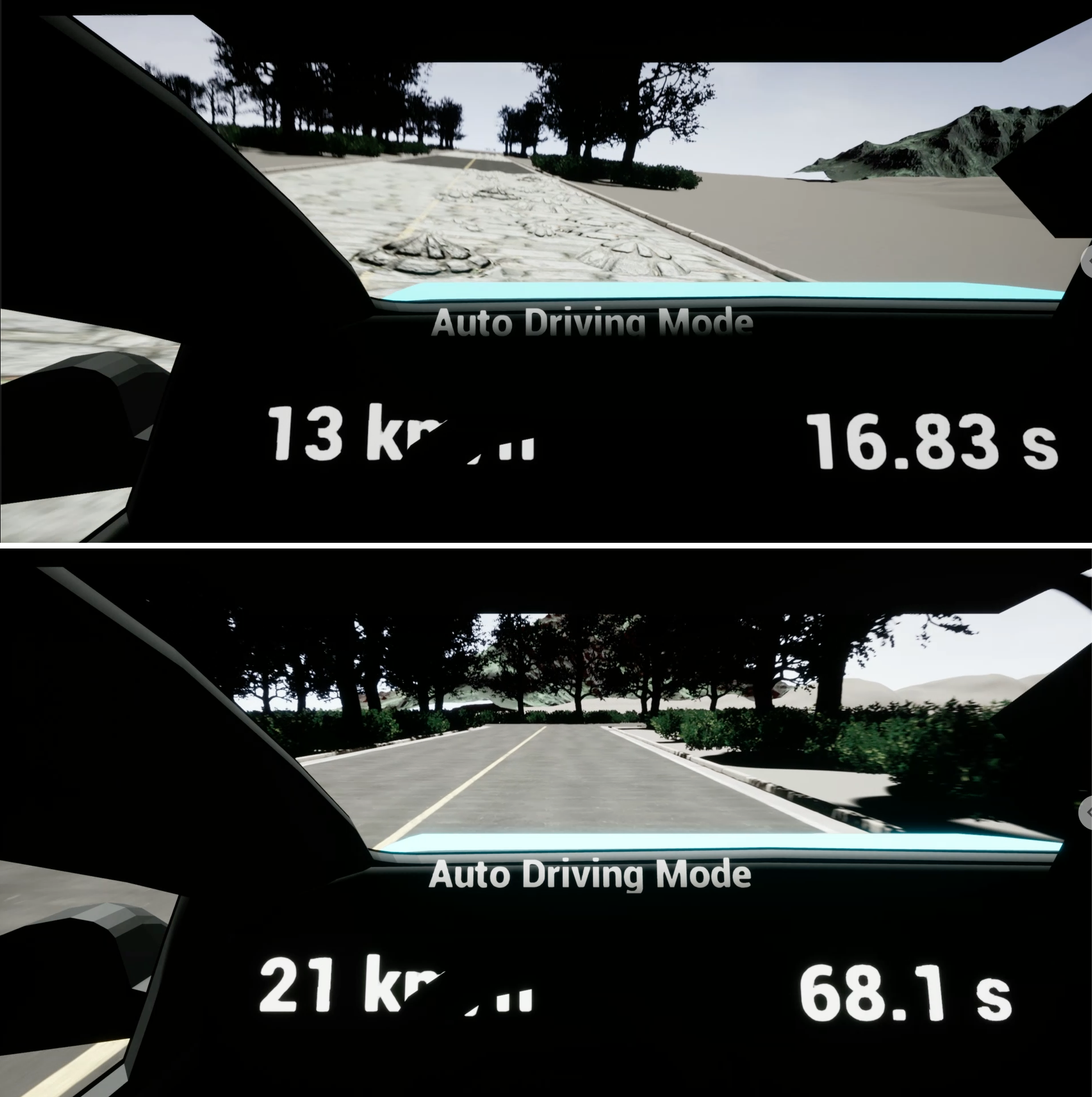}
    }
    \vspace{-0.35cm}
  \caption{The Motivating Scenario and User Study built on Unreal Engine.}
    \label{scenariomap}
    \vspace{-0.3cm}
\end{figure}

To illustrate the preference adaptation framework, we focus on an autonomous driving system's route choice scenario. Fig.~\ref{scenariomap}(a) shows an autonomous vehicle transport the user from a start to end point, with multiple options: 1. The Shortest Route: Located at the bottom, it features a rough stone road with considerable noise; 2. The Middle Route: This path offers a fine stone road bordered by bushes, but is not immune to noise disturbances; 3. The Third Route: Predominantly consists of a fine stone road; 
4. The Scenic Route: The longest and most winding, it is enveloped by tree canopies, offering a smooth and flat ride within a tranquil and scenic setting.


Key quality attributes for this route selection include: \emph{Efficiency}: the vehicle's travel time.  \emph{Aesthetic Appeal}: the visual and auditory impact of the surroundings, with factors like trees enhancing aesthetic appeal and ambient noise reducing it. \emph{Road Condition}: the quality of the road infrastructure and its maintenance, significantly influencing ride smoothness. Notably, the nature of the terrain, such as a bumpy road, can greatly impact passenger comfort~\cite{routechoice}\footnote{Autonomous driving scenarios span a broad spectrum of attributes, including safety, energy consumption, interaction with other vehicles, and more. However, to illustrate the framework and user study later effectively, we narrowed our focus to three attributes. This decision was guided by two key factors. Firstly, these attributes are directly perceptible. In user studies, participants can quickly discern the implications of travel time (Efficiency), visual surroundings (Aesthetic Appeal), and ride comfort (Road Condition).  Second, they are easily simulated in Unreal Engine and quantifiable, allowing clear representation and measurement of user experience in our study. 
While crucial, attributes like energy consumption aren't as immediately noticeable in short user studies. In contrast, the impact of picturesque routes or the feel of a bumpy road is immediately evident. }.

The vehicle selects routes based on distinct user preferences. For example, a preference weighting of $\langle$ 0.333 (Efficiency), 0.333 (Aesthetic Appeal), 0.333 (Road Condition) $\rangle$ indicates an equal emphasis on three attributes. A certain route is evaluated by aggregating the utility values associated with each quality attribute,  defined as $U(route) = \sum\limits_{i=1}^{n} w_i \times u_i$ where $w_i$ stands for the preference value assigned to a particular quality attribute $i$ and $u_i$ denotes the utility value for that attribute. 
For \emph{Efficiency}, the utility is based on travel distance, assessed by the number of segments traversed. In the case of \emph{Road Condition}, rougher roads are assigned lower utility due to increased discomfort, while smoother roads have higher utility indicating a more comfortable and less wearing journey. Similarly, \emph{Aesthetic Appeal} is evaluated based on environmental factors like noise levels and greenery. Areas with less aesthetic appeal, such as noisy or treeless zones, incur lower utility, while more serene and green environments have higher utility. 
The optimal route is determined by maximizing the aggregation utility function, which can be implemented through various algorithms like A star algorithm. In this typical scenario as in Fig.~\ref{scenariomap}-(a), the best choice is route 3 for the given user preference. 
Details regarding utility assignments for different quality attributes and the decision-making process for selecting the optimal route based on user preferences are not the main focus of this paper. Those looking for a deeper understanding of the underlying calculations can refer to the work~\cite{nianyuPreference}.

In our research, preference adaptation transforms into an optimization problem seeking ideal weight configurations by iteratively refining attribute weights in light of user feedback. We aim to achieve three key objectives through these updates: (1) Preference Divergence, aiming to minimize the variance between initial and updated preferences while ensuring user satisfaction, based on the premise that preference shifts occur incrementally over time \cite{driverPreference}; (2) Complaint Avoidance, ensuring that the optimal trajectory derived from updated preferences does not intersect with states previously identified as problematic by users; and (3) Implicit Constraints,  which involves maintaining specific hierarchical relationships within the attribute weights, ensuring that certain attributes are prioritized based on user feedback such as "excessive noise!" or "excessive bumpiness!". These objectives are encapsulated within a unified fitness function encoding user complaints, formulated as $f(p)=\lambda_1f_1+\lambda_2f_2+\lambda_3f_3$, where $\lambda_1$, $\lambda_2$, $\lambda_3$ are positive coefficients, $f_1$, $f_2$ and $f_3$ represent the objectives mentioned, and $p=\langle w_1,...,w_n\rangle$ denotes the preference vector. 


To align user preferences with these objectives, the approach employs a Genetic Algorithm (GA) framework, renowned for its efficacy in solving such search problems. Within this framework, the fitness function corresponds to the objective function described earlier. Each attribute weight is conceptualized as an individual within the GA population, represented as vectors like $p_1=\langle w_1,w_2,...,w_n \rangle$ and $p_2=\langle w_1',w_2',...,w_n' \rangle$. In this context, each weight, $w_i$, functions as a gene within these individuals. The crossover operation in GA entails exchanging a subset of weights between individuals, generating novel combinations. Post-crossover, individuals exceeding the total weight constraint are either normalized or discarded to maintain validity. The mutation operation involves minor, random adjustments to a weight, with compensatory changes in other weights to preserve the sum constraint. The specifics of these GA operations are further elucidated in the appendix. By employing GA, we establish a systematic, evolutionary method to iteratively refine and optimize preference vectors, adhering to the predefined objectives and user feedback.

\section{User Study}
\label{sec:userstudy}

To empirically assess the efficacy of the proposed preference adaptation framework with user complaints, our user study is designed to answer the following research questions. Informed consent forms were distributed prior to the participation in the experiment, and data were processed in line with the recommendations of the ethics board of the university of researchers responsible for the study.   
\begin{itemize}
\item RQ1: Quantitative Analysis of Preference Alignment. Does the framework, employing a complaints feedback mechanism and genetic algorithm for preference adaptation, effectively update preferences to align more closely with users' self-reported preferences? 
\item RQ2: Qualitative Assessment of User Satisfaction. How does the integration of the framework, particularly its adaptation of preferences based on user complaints, impact user satisfaction with system behaviors?
\end{itemize}

\subsection{Procedure}

Our user study commenced with a baseline assumption: all users have equal preference values across three quality attributes. These initial settings were considered as starting or 'outdated' preferences (i.e., $\langle$0.333, 0.333, 0.334$\rangle$). Utilizing Unreal Engine~\cite{UnrealEngine2024}, a 3D computer graphics game engine, we developed a series of maps, as shown in Fig.~\ref{scenariomap}, as the experimental environment. 
As the study progressed, we focused on dynamically updating these preferences by inferring from each user's complaints regarding the routes recommended by the system. This adaptation process is a practical application of 'dynamic updating' of preferences, illustrating the system's ability to evolve and align with real-time changes in user preferences. The study was structured into three sequential phases: a pre-experiment questionnaire, the main experiment, and a post-experiment questionnaire complemented by an interview.

\textbf{Pre-experiment Questionnaire.}
Before delving into the main experiment, the session began with an interactive Q\&A segment managed by the organizer. Participants were tasked with filling out an initial questionnaire, encompassing: (1) essential personal and demographic data; 
and (2) a self-reported preference after viewing a sample video. This video depicted a driver's first-person perspective under various road environments, such as routes characterized by trees, bushes, fine stones, rough stones, and noise. These preferences were expected to remain consistent throughout the duration of the study, serving as a benchmark against which any shifts in preferences could be measured and analyzed.


\textbf{Main Experiment.}
Within the UE simulator, we designed three tailored maps, with each map comprising four distinct driving routes similar to Fig.~\ref{scenariomap}-(a). For each of these maps, participants underwent a structured set of tasks:
(1) Viewing the Algorithm-Recommended Route: 
Initiated by the organizer, participants watched a screen-shared video, representing an autonomous-driven journey along the route recommended by the system. 
Recommendations for subsequent maps were modified, taking into account feedback from the previous map's experience. 
(2) Complaints and Route Scoring: After viewing the suggested route, participants were given the opportunity to voice complaints of general discontent and 
specific discontent. Available options encompassed: dislike of the road, excessive noise, excessive bumpiness, excessive distance, or no complaints. 
We focused on these complaint types as they precisely capture users' dissatisfaction stemming from system behavior, providing a clearer insight into explicit user intentions over preference-based grievances. Additionally, based on their experience with the route, they were requested to quantify their experience, rating their satisfaction on a standardized scale from very dissatisfied to very satisfied. 
(3) Viewing and Scoring Alternative Routes: Subsequently, participants were shown videos of the other three available routes within that particular map. After each video, they were instructed to provide a rating, reflecting their satisfaction with the route, using the same scale.


\textbf{Post-experiment Questionnaire and Interview.}
Upon completion of the primary tasks, participants were guided to a concluding assessment phase, which included a questionnaire and supplementary interviews.
This assessment encapsulated: (1) a refreshed self-report on their preferences after interacting with the simulation; (2) an opportunity to provide feedback or complaints about the route calculated based on their initial self-reported preferences in the third map, especially if it differed from the route suggested by the system after updating preferences with user feedback; 
(3) open-ended questions designed to delve deeper into their experiences and collect personalized and detailed feedback; 
and (4) a targeted interview for select participants, specifically those who demonstrated significant variances between their pre and post-experiment self-reported preferences were engaged in individual interviews to better understand their perspectives and rationale.

\subsection{Participants} We recruited 20 participants, comprising 12 males and 8 females, from university campus. Each participant is presented with high-definition (1080p) video portrayals of the autonomous driving experience, generated from the UE simulator (Refer to Fig.~\ref{scenariomap}-(b)) . 
All participants possessed normal hearing and either normal vision or vision corrected to normal standards. Participant ages spanned from 19 to 49 years, with a mean age of 27.5 and a standard deviation of 4.1. Participation was on a voluntary basis, and no financial incentives were offered.


\subsection{Metrics and Data Processing}
For clarity, we delineate the metrics associated with RQ1 and RQ2, along with the associated data collection and processing methods.

\textbf{RQ1}: Addressing RQ1, our primary metric centers on evaluating the preference similarity. 
This involves a detailed comparison between the user preferences dynamically refined by our algorithm and the users' self-reported preferences as gathered from the post-experiment questionnaire. Initially, each participant's preference is set to a default vector of $\langle 0.333, 0.333, 0.334 \rangle$. This preference vector undergoes sequential adaptations after each map experiment, contingent upon the first, second, and third instances of user complaints, if any. To quantitatively assess these preference similarities, we represent the preferences as three-dimensional vectors and employ Cosine similarity as our metric for evaluation.


\textbf{RQ2}: For RQ2, we considered three key metrics:
(1) Frequency of complaints: The recurrence of user complaints serves as an insightful indicator of user satisfaction with the system's behaviors. We counted the number of user complaints about the recommended routes across the three maps;
(2) Self-assessed satisfaction: To glean insights into user satisfaction, participants were presented with a Likert scale query in the post-experiment questionnaire: "Do you agree: The system-recommended routes increasingly satisfy you?" The response spectrum was articulated on a five-point Likert scale, ranging from "Strongly Disagree" to "Strongly Agree;"
(3) Scoring and ranking of recommended routes: As part of the experimental procedure, participants were prompted to score each route on a Likert scale from 1 to 5 post-viewing,  with the scale defined as follows: 1 - Very Dissatisfied, 2 - Dissatisfied, 3 - Neutral, 4 - Satisfied, 5 - Very Satisfied. This facilitated the extraction of both the rank (relative to the other three routes within the same map) and numerical evaluation for the recommended routes. To ensure comparability across diverse scenarios, user satisfaction values for each map were subjected to min-max normalization.

\section{Results and Analysis}
\label{sec:results}
\begin{figure}[h!tbp]
    \vspace{-0.2cm}
    \centering
    \subfigure[Temporal evolution of preference similarity.]{
        \includegraphics[width=0.47\textwidth]{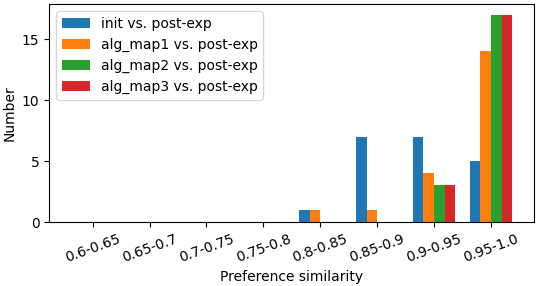}
    }
    \quad 
    \subfigure[Distribution of preference similarity values.]{
    \includegraphics[width=0.47\textwidth]{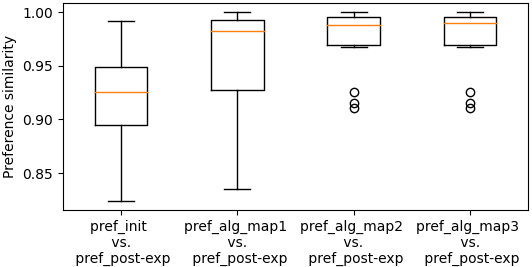}
    }
    \quad 
    \subfigure[Route satisfaction ranking per map.]{
        \includegraphics[width=0.47\textwidth]{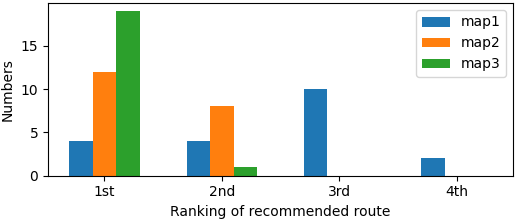}
    }
    \subfigure[Distribution of user satisfaction scores for recommended routes.]{
        \includegraphics[width=0.48\textwidth]{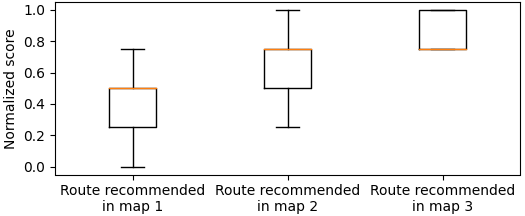}
    }
    \vspace{-0.2cm}
    \caption{User Study Results}
    \label{fig: RQ1}
    \vspace{-0.3cm}

\end{figure}

Fig.\ref{fig: RQ1}.(a) and (b) trace the temporal evolution of preference similarity.
Fig.\ref{fig: RQ1}.(a) delineates the distribution of these similarities. The median values for these similarities are reported as 0.926 (with an IQR of 0.949 - 0.895), 0.982 (IQR: 0.993 - 0.927), 0.988 (IQR: 0.995 - 0.970), and 0.990 (IQR: 0.995 - 0.970). 
This pattern signifies a notable trend: the initial balanced setting of preference weights contributed to the initial similarity score (0.926), as it closely aligns with the general tendency of users to distribute their preferences evenly across multiple attributes. As user complaints accumulate, there is a progressive alignment between the algorithmically updated preferences and the users' self-reported preferences, pushing the Cosine similarity metric closer to 1. 
It is also crucial to highlight the presence of three outliers within this data. Remarkably, in two of these outlier instances, the preferences updated by the algorithm coincide with the route recommendations as reported by the users in their self-reported preferences. This observation underscores the nuanced efficacy of the algorithm in adapting to user feedback and aligning with user preferences over time.

In our analysis of user complaints across the three maps, we observed a decreasing trend: 20 complaints for Map 1, 11 for Map 2, and only 2 for Map 3. This reduction in complaints can be attributed to the preference adaptation. After addressing the initial set of complaints from Map 1, the algorithm adjusted the recommended routes, which resulted in a significant drop in user dissatisfaction, as evidenced by the decrease in complaints from 20 to 11. Further refinements based on subsequent feedback led to an even more pronounced decline in complaints, down to just 2 in Map 3. 
Regarding participant satisfaction with the system recommendation, the response was overwhelmingly positive. Out of the participants, 12 expressed ``Strongly Agree'', and 8 chose ``Agree'' when asked ``if the system-recommended routes increasingly met their expectations''. Notably, there were no neutral or negative responses (``Neutral'', ``Disagree'', or ``Strongly Disagree'').

Fig.\ref{fig: RQ1}.(c) and (d) offer insights into the satisfaction ranking and scores of the recommended routes. 
As depicted in Fig.~\ref{fig: RQ1}.(c), for map 1, based on the algorithm's initial default preferences, the route rankings were as follows: 4 participants ranked it first, 4 participants second, 10 participants third, and 2 participants fourth. 
However, subsequent maps showed significant improvements; in map 2, 12 participants ranked the recommended route first, and 8 participants second. 
In map 3, 19 participants ranked the recommended route first, and 1 participant second. 
This progression highlights the effectiveness of preference adaptation, demonstrating that the system-recommended routes increasingly aligned with users’ best options.
Delving into the scores shown in Fig.~\ref{fig: RQ1}.(d), a notable increase in median satisfaction scores is observed across the maps: from 0.5 (IQR: 0.5 - 0.35) in map 1, the score rose to 0.75 (IQR: 0.75 - 0.5) for map 2, and further to 0.75 (IQR: 1 - 0.75) for map 3. 
This trend underscores that with an increasing number of complaints, users' scoring of the system's recommended routes progressively improves.




The results of our study underscore the importance of considering the dynamic nature of user preferences in the development of user-centric systems. Specifically, they validate the effectiveness of our GA-based preference adaptation, which adeptly integrates a feedback mechanism via user complaints. While our approach uses self-reported preferences as a comparative benchmark and the results indicate a gradual convergence of the algorithmically updated preferences towards these self-reported preferences, it is crucial to note that relying solely on direct user input for setting preferences may not be entirely effective, especially in real-world scenarios. This is supported by the following considerations: 


\begin{itemize}
\item Inaccuracy and Contradictions in Self-Reported Preferences:
Our analysis revealed that users might not always provide an accurate reflection of their true preferences, which can sometimes be confusing or contradictory.  This was apparent when we examined the route choices in Map 3. We compared routes based on two different sources: participants' self-reported preferences and those generated by our algorithm. Remarkably, in 85\% (17 out of 20) of cases, the routes suggested by both methods were consistent. However, among the remaining three participants, one found equal satisfaction with both routes, indicating potential inconsistencies in their self-reported preferences. The second participant showed a clear preference for the algorithm-suggested route, while the third participant voiced complaints about the self-reported preference route. These observations align with findings that humans often struggle to articulate their objectives accurately \cite{seamskenote2021, articleneuroscience, DBLP:conf/dagstuhl/2009adaptive}. As such, the gradual reduction in user complaints over time becomes a more reliable and complementary indicator of the system's effectiveness in aligning with users' evolving preferences.

\item Challenges in Timely Setting of User Preferences :
Determining the optimal moments to solicit updated preferences from users poses a significant challenge, given the fluidity of user preferences. Our pre- and post-experiment analysis of self-reported preferences revealed that only 35\% (7 out of 20) of participants retained identical preferences (as indicated by a cosine similarity of 1). Notably, 25\% (5 out of 20) exhibited a similarity score of less than 0.9, with one participant's score falling below 0.5. A post-experiment interview with this participant highlighted the reasons behind this significant shift. Initially, the participant was averse to longer routes. However, after viewing videos during the experiment, which depicted high levels of noise and bumpiness, the participant's preference changed due to the discomfort these factors caused.

\end{itemize}

\section{Conclusion and Future Work}
\label{sec:conclusion}

To tackle the complexity and dynamism of user preferences, our study introduces a novel framework underpinned by the "human-on-the-loop" concept. This framework enables the system to adapt to human preferences using a genetic algorithm driven by user dissatisfaction, thereby aligning system behavior more closely with human expectations. While the preference adaptation in our user study of autonomous driving scenario —transitioning from initial settings to more closely match human self-reported preferences—is reactive in nature, the results demonstrate its quantitative effectiveness in aligning preferences and qualitative enhancement in user satisfaction. 


In our study, conducted within the Unreal Engine environment, we effectively simulated three key quality attributes relevant to route choice scenarios. The aim of selecting these particular attributes was to ensure that our study results are coherent and rooted in the tangible experiences of participants. Although this was instrumental in achieving our immediate research goals, we acknowledge the necessity of including a broader range of attributes in future studies to gain a more comprehensive understanding of user preferences in autonomous driving contexts. 
Besides, a notable limitation concerns the perception and impact of Efficiency. 
The duration of each route in the simulation ranged from 40 to 80 seconds. This narrow time range, coupled with the absence of significant consequences for longer travel times, may have inadequately captured the practical implications of efficiency in route selection. Consequently, this limitation might have influenced the participants' ability to discern and prioritize efficiency as a critical attribute, posing a threat to the validity of our findings regarding efficiency preferences. 


To optimize our framework and further align it with user needs, future enhancements could focus on implementing more refined feedback mechanisms. A key area of improvement involves developing a system capable of discerning varying degrees of user concerns. For example, differentiating between "slightly noisy" and "extremely noisy" environments would offer a more detailed understanding of user preferences, thereby enabling a more nuanced response to their specific complaints. Additionally, the integration of large language models holds promise for augmenting our framework's capacity to parse and analyze user feedback. These models are particularly adept at processing natural language, which ranges from specific complaints to general comments. Their inclusion would enable our system to more effectively adapt the fitness function within the genetic algorithm, ensuring that it responds more precisely to the issues and preferences identified in user feedback.

\bibliographystyle{ACM-Reference-Format}
\bibliography{sample-base}


\begin{thebibliography}{25}


\ifx \showCODEN    \undefined \def \showCODEN     #1{\unskip}     \fi
\ifx \showDOI      \undefined \def \showDOI       #1{#1}\fi
\ifx \showISBNx    \undefined \def \showISBNx     #1{\unskip}     \fi
\ifx \showISBNxiii \undefined \def \showISBNxiii  #1{\unskip}     \fi
\ifx \showISSN     \undefined \def \showISSN      #1{\unskip}     \fi
\ifx \showLCCN     \undefined \def \showLCCN      #1{\unskip}     \fi
\ifx \shownote     \undefined \def \shownote      #1{#1}          \fi
\ifx \showarticletitle \undefined \def \showarticletitle #1{#1}   \fi
\ifx \showURL      \undefined \def \showURL       {\relax}        \fi
\providecommand\bibfield[2]{#2}
\providecommand\bibinfo[2]{#2}
\providecommand\natexlab[1]{#1}
\providecommand\showeprint[2][]{arXiv:#2}

\bibitem[Abe et~al\mbox{.}(2024)]%
        {Abe2024Enhancing}
\bibfield{author}{\bibinfo{person}{Ryotaro Abe}, \bibinfo{person}{Jinyu Cai}, \bibinfo{person}{Tianchen Wang}, \bibinfo{person}{Jialong Li}, \bibinfo{person}{Shinichi Honiden}, {and} \bibinfo{person}{Kenji Tei}.} \bibinfo{year}{2024}\natexlab{}.
\newblock \showarticletitle{Towards Enhancing Driver's Perceived Safety in Autonomous Driving: A Shield-based Approach}. In \bibinfo{booktitle}{\emph{Intelligent Systems Design and Applications}}. \bibinfo{publisher}{Springer}.
\newblock


\bibitem[Cheng et~al\mbox{.}(2009)]%
        {DBLP:conf/dagstuhl/2009adaptive}
\bibfield{editor}{\bibinfo{person}{Betty H.~C. Cheng}, \bibinfo{person}{Rog{\'{e}}rio de Lemos}, \bibinfo{person}{Holger Giese}, \bibinfo{person}{Paola Inverardi}, {and} \bibinfo{person}{Jeff Magee}} (Eds.). \bibinfo{year}{2009}\natexlab{}.
\newblock \bibinfo{booktitle}{\emph{Software Engineering for Self-Adaptive Systems [outcome of a Dagstuhl Seminar]}}. \bibinfo{series}{Lecture Notes in Computer Science}, Vol.~\bibinfo{volume}{5525}. \bibinfo{publisher}{Springer}.
\newblock


\bibitem[Chu et~al\mbox{.}(2023)]%
        {DBLP:conf/chi/ChuZSGLGDZ23}
\bibfield{author}{\bibinfo{person}{Mengdi Chu}, \bibinfo{person}{Keyu Zong}, \bibinfo{person}{Xin Shu}, \bibinfo{person}{Jiangtao Gong}, \bibinfo{person}{Zhicong Lu}, \bibinfo{person}{Kaimin Guo}, \bibinfo{person}{Xinyi Dai}, {and} \bibinfo{person}{Guyue Zhou}.} \bibinfo{year}{2023}\natexlab{}.
\newblock \showarticletitle{Work with {AI} and Work for {AI:} Autonomous Vehicle Safety Drivers' Lived Experiences}. In \bibinfo{booktitle}{\emph{Proceedings of the 2023 {CHI} Conference on Human Factors in Computing Systems, {CHI} 2023, Hamburg, Germany, April 23-28, 2023}}, \bibfield{editor}{\bibinfo{person}{Albrecht Schmidt}, \bibinfo{person}{Kaisa V{\"{a}}{\"{a}}n{\"{a}}nen}, \bibinfo{person}{Tesh Goyal}, \bibinfo{person}{Per~Ola Kristensson}, \bibinfo{person}{Anicia Peters}, \bibinfo{person}{Stefanie Mueller}, \bibinfo{person}{Julie~R. Williamson}, {and} \bibinfo{person}{Max~L. Wilson}} (Eds.). \bibinfo{publisher}{{ACM}}, \bibinfo{pages}{753:1--753:16}.
\newblock


\bibitem[Dillen et~al\mbox{.}(2020a)]%
        {passengercomfortAnxiety}
\bibfield{author}{\bibinfo{person}{Nicole Dillen}, \bibinfo{person}{Marko Ilievski}, \bibinfo{person}{Edith Law}, \bibinfo{person}{Lennart~E. Nacke}, \bibinfo{person}{Krzysztof Czarnecki}, {and} \bibinfo{person}{Oliver Schneider}.} \bibinfo{year}{2020}\natexlab{a}.
\newblock \showarticletitle{Keep Calm and Ride Along: Passenger Comfort and Anxiety as Physiological Responses to Autonomous Driving Styles}. In \bibinfo{booktitle}{\emph{{CHI} '20: {CHI} Conference on Human Factors in Computing Systems, Honolulu, HI, USA, April 25-30, 2020}}, \bibfield{editor}{\bibinfo{person}{Regina Bernhaupt}, \bibinfo{person}{Florian~'Floyd' Mueller}, \bibinfo{person}{David Verweij}, \bibinfo{person}{Josh Andres}, \bibinfo{person}{Joanna McGrenere}, \bibinfo{person}{Andy Cockburn}, \bibinfo{person}{Ignacio Avellino}, \bibinfo{person}{Alix Goguey}, \bibinfo{person}{Pernille Bj{\o}n}, \bibinfo{person}{Shengdong Zhao}, \bibinfo{person}{Briane~Paul Samson}, {and} \bibinfo{person}{Rafal Kocielnik}} (Eds.). \bibinfo{publisher}{{ACM}}, \bibinfo{pages}{1--13}.
\newblock
\urldef\tempurl%
\url{https://doi.org/10.1145/3313831.3376247}
\showDOI{\tempurl}


\bibitem[Dillen et~al\mbox{.}(2020b)]%
        {DBLP:conf/chi/DillenILNCS20}
\bibfield{author}{\bibinfo{person}{Nicole Dillen}, \bibinfo{person}{Marko Ilievski}, \bibinfo{person}{Edith Law}, \bibinfo{person}{Lennart~E. Nacke}, \bibinfo{person}{Krzysztof Czarnecki}, {and} \bibinfo{person}{Oliver Schneider}.} \bibinfo{year}{2020}\natexlab{b}.
\newblock \showarticletitle{Keep Calm and Ride Along: Passenger Comfort and Anxiety as Physiological Responses to Autonomous Driving Styles}. In \bibinfo{booktitle}{\emph{{CHI} '20: {CHI} Conference on Human Factors in Computing Systems, Honolulu, HI, USA, April 25-30, 2020}}, \bibfield{editor}{\bibinfo{person}{Regina Bernhaupt}, \bibinfo{person}{Florian~'Floyd' Mueller}, \bibinfo{person}{David Verweij}, \bibinfo{person}{Josh Andres}, \bibinfo{person}{Joanna McGrenere}, \bibinfo{person}{Andy Cockburn}, \bibinfo{person}{Ignacio Avellino}, \bibinfo{person}{Alix Goguey}, \bibinfo{person}{Pernille Bj{\o}n}, \bibinfo{person}{Shengdong Zhao}, \bibinfo{person}{Briane~Paul Samson}, {and} \bibinfo{person}{Rafal Kocielnik}} (Eds.). \bibinfo{publisher}{{ACM}}, \bibinfo{pages}{1--13}.
\newblock


\bibitem[{Epic Games}(2024)]%
        {UnrealEngine2024}
\bibfield{author}{\bibinfo{person}{{Epic Games}}.} \bibinfo{year}{2024}\natexlab{}.
\newblock \bibinfo{title}{Unreal Engine - The Most Powerful Real-Time 3D Creation Tool}.
\newblock
\newblock
\urldef\tempurl%
\url{https://www.unrealengine.com/en-US}
\showURL{%
\tempurl}
\newblock
\shownote{Accessed: 2024-01-23}.


\bibitem[Gouveia and Epstein(2023)]%
        {preferencePersonalTracking}
\bibfield{author}{\bibinfo{person}{R{\'{u}}ben Gouveia} {and} \bibinfo{person}{Daniel~A. Epstein}.} \bibinfo{year}{2023}\natexlab{}.
\newblock \showarticletitle{This Watchface Fits with my Tattoos: Investigating Customisation Needs and Preferences in Personal Tracking}. In \bibinfo{booktitle}{\emph{Proceedings of the 2023 {CHI} Conference on Human Factors in Computing Systems, {CHI} 2023, Hamburg, Germany, April 23-28, 2023}}, \bibfield{editor}{\bibinfo{person}{Albrecht Schmidt}, \bibinfo{person}{Kaisa V{\"{a}}{\"{a}}n{\"{a}}nen}, \bibinfo{person}{Tesh Goyal}, \bibinfo{person}{Per~Ola Kristensson}, \bibinfo{person}{Anicia Peters}, \bibinfo{person}{Stefanie Mueller}, \bibinfo{person}{Julie~R. Williamson}, {and} \bibinfo{person}{Max~L. Wilson}} (Eds.). \bibinfo{publisher}{{ACM}}, \bibinfo{pages}{327:1--327:15}.
\newblock
\urldef\tempurl%
\url{https://doi.org/10.1145/3544548.3580955}
\showDOI{\tempurl}


\bibitem[Gratzl et~al\mbox{.}(2013)]%
        {DBLP:journals/tvcg/GratzlLGPS13}
\bibfield{author}{\bibinfo{person}{Samuel Gratzl}, \bibinfo{person}{Alexander Lex}, \bibinfo{person}{Nils Gehlenborg}, \bibinfo{person}{Hanspeter Pfister}, {and} \bibinfo{person}{Marc Streit}.} \bibinfo{year}{2013}\natexlab{}.
\newblock \showarticletitle{LineUp: Visual Analysis of Multi-Attribute Rankings}.
\newblock \bibinfo{journal}{\emph{{IEEE} Trans. Vis. Comput. Graph.}} \bibinfo{volume}{19}, \bibinfo{number}{12} (\bibinfo{year}{2013}), \bibinfo{pages}{2277--2286}.
\newblock


\bibitem[Karagiannakis et~al\mbox{.}(2014)]%
        {articleneuroscience}
\bibfield{author}{\bibinfo{person}{Giannis Karagiannakis}, \bibinfo{person}{Anna Baccaglini-Frank}, {and} \bibinfo{person}{Yiannis Papadatos}.} \bibinfo{year}{2014}\natexlab{}.
\newblock \showarticletitle{Mathematical learning difficulties subtypes classification}.
\newblock \bibinfo{journal}{\emph{Frontiers in Human Neuroscience}}  \bibinfo{volume}{8} (\bibinfo{date}{01} \bibinfo{year}{2014}).
\newblock


\bibitem[Kephart(2021)]%
        {seamskenote2021}
\bibfield{author}{\bibinfo{person}{Jeffrey Kephart}.} \bibinfo{year}{2021}\natexlab{}.
\newblock \showarticletitle{Viewing Autonomic Computing through the Lens of Embodied Artificial Intelligence: A Self-Debate.}
\newblock \bibinfo{journal}{\emph{Keynote at the 16th Symposium on Software Engineering for Adaptive and Self-Managing Systems. (SEAMS 2021)}} (\bibinfo{year}{2021}).
\newblock


\bibitem[Kim et~al\mbox{.}(2021)]%
        {kim2021guiding}
\bibfield{author}{\bibinfo{person}{Keunwoo Kim}, \bibinfo{person}{Minjung Park}, {and} \bibinfo{person}{Youn-kyung Lim}.} \bibinfo{year}{2021}\natexlab{}.
\newblock \showarticletitle{Guiding preferred driving style using voice in autonomous vehicles: An on-road wizard-of-oz study}. In \bibinfo{booktitle}{\emph{Designing Interactive Systems Conference 2021}}. \bibinfo{pages}{352--364}.
\newblock


\bibitem[Kuhlman et~al\mbox{.}(2018)]%
        {DBLP:conf/cikm/KuhlmanVDNDPRH18}
\bibfield{author}{\bibinfo{person}{Caitlin Kuhlman}, \bibinfo{person}{MaryAnn~Van Valkenburg}, \bibinfo{person}{Diana Doherty}, \bibinfo{person}{Malika Nurbekova}, \bibinfo{person}{Goutham Deva}, \bibinfo{person}{Zarni Phyo}, \bibinfo{person}{Elke~A. Rundensteiner}, {and} \bibinfo{person}{Lane Harrison}.} \bibinfo{year}{2018}\natexlab{}.
\newblock \showarticletitle{Preference-driven Interactive Ranking System for Personalized Decision Support}. In \bibinfo{booktitle}{\emph{Proceedings of the 27th {ACM} International Conference on Information and Knowledge Management, {CIKM} 2018, Torino, Italy, October 22-26, 2018}}, \bibfield{editor}{\bibinfo{person}{Alfredo Cuzzocrea}, \bibinfo{person}{James Allan}, \bibinfo{person}{Norman~W. Paton}, \bibinfo{person}{Divesh Srivastava}, \bibinfo{person}{Rakesh Agrawal}, \bibinfo{person}{Andrei~Z. Broder}, \bibinfo{person}{Mohammed~J. Zaki}, \bibinfo{person}{K.~Sel{\c{c}}uk Candan}, \bibinfo{person}{Alexandros Labrinidis}, \bibinfo{person}{Assaf Schuster}, {and} \bibinfo{person}{Haixun Wang}} (Eds.). \bibinfo{publisher}{{ACM}}, \bibinfo{pages}{1931--1934}.
\newblock


\bibitem[Lesch et~al\mbox{.}(2021)]%
        {percomworkshoppreference}
\bibfield{author}{\bibinfo{person}{Veronika Lesch}, \bibinfo{person}{Marius Hadry}, \bibinfo{person}{Samuel Kounev}, {and} \bibinfo{person}{Christian Krupitzer}.} \bibinfo{year}{2021}\natexlab{}.
\newblock \showarticletitle{Utility-based Vehicle Routing Integrating User Preferences}. In \bibinfo{booktitle}{\emph{19th {IEEE} International Conference on Pervasive Computing and Communications Workshops and other Affiliated Events, PerCom Workshops 2021, Kassel, Germany, March 22-26, 2021}}. \bibinfo{publisher}{{IEEE}}, \bibinfo{pages}{263--268}.
\newblock


\bibitem[Li et~al\mbox{.}(2021)]%
        {9391909}
\bibfield{author}{\bibinfo{person}{Jialong Li}, \bibinfo{person}{Zhenyu Mao}, \bibinfo{person}{Zhen Cao}, \bibinfo{person}{Kenji Tei}, {and} \bibinfo{person}{Shinichi Honiden}.} \bibinfo{year}{2021}\natexlab{}.
\newblock \showarticletitle{Self-adaptive Hydroponics Care System for Human-hydroponics Coexistence}. In \bibinfo{booktitle}{\emph{2021 IEEE 3rd Global Conference on Life Sciences and Technologies (LifeTech)}}. \bibinfo{pages}{204--206}.
\newblock


\bibitem[Li et~al\mbox{.}(2023)]%
        {nianyuPreference}
\bibfield{author}{\bibinfo{person}{Nianyu Li}, \bibinfo{person}{Mingyue Zhang}, \bibinfo{person}{Jialong Li}, \bibinfo{person}{Eunsuk Kang}, {and} \bibinfo{person}{Kenji Tei}.} \bibinfo{year}{2023}\natexlab{}.
\newblock \showarticletitle{Preference Adaptation: user satisfaction is all you need!}. In \bibinfo{booktitle}{\emph{18th {IEEE/ACM} Symposium on Software Engineering for Adaptive and Self-Managing Systems, {SEAMS} 2023, Melbourne, Australia, May 15-16, 2023}}. \bibinfo{publisher}{{IEEE}}, \bibinfo{pages}{133--144}.
\newblock


\bibitem[Ling et~al\mbox{.}(2021)]%
        {9723811}
\bibfield{author}{\bibinfo{person}{Jiali Ling}, \bibinfo{person}{Jialong Li}, \bibinfo{person}{Kenji Tei}, {and} \bibinfo{person}{Shinichi Honiden}.} \bibinfo{year}{2021}\natexlab{}.
\newblock \showarticletitle{Towards Personalized Autonomous Driving: An Emotion Preference Style Adaptation Framework}. In \bibinfo{booktitle}{\emph{2021 IEEE International Conference on Agents (ICA)}}. \bibinfo{pages}{47--52}.
\newblock
\urldef\tempurl%
\url{https://doi.org/10.1109/ICA54137.2021.00015}
\showDOI{\tempurl}


\bibitem[Pajer et~al\mbox{.}(2017)]%
        {DBLP:journals/tvcg/PajerSTSMP17}
\bibfield{author}{\bibinfo{person}{Stephan Pajer}, \bibinfo{person}{Marc Streit}, \bibinfo{person}{Thomas Torsney{-}Weir}, \bibinfo{person}{Florian Spechtenhauser}, \bibinfo{person}{Torsten M{\"{o}}ller}, {and} \bibinfo{person}{Harald Piringer}.} \bibinfo{year}{2017}\natexlab{}.
\newblock \showarticletitle{WeightLifter: Visual Weight Space Exploration for Multi-Criteria Decision Making}.
\newblock \bibinfo{journal}{\emph{{IEEE} Trans. Vis. Comput. Graph.}} \bibinfo{volume}{23}, \bibinfo{number}{1} (\bibinfo{year}{2017}), \bibinfo{pages}{611--620}.
\newblock
\urldef\tempurl%
\url{https://doi.org/10.1109/TVCG.2016.2598589}
\showURL{%
\tempurl}


\bibitem[Pan et~al\mbox{.}(2020)]%
        {driverPreference}
\bibfield{author}{\bibinfo{person}{Menghai Pan}, \bibinfo{person}{Weixiao Huang}, \bibinfo{person}{Yanhua Li}, \bibinfo{person}{Xun Zhou}, \bibinfo{person}{Zhenming Liu}, \bibinfo{person}{Rui Song}, \bibinfo{person}{Hui Lu}, \bibinfo{person}{Zhihong Tian}, {and} \bibinfo{person}{Jun Luo}.} \bibinfo{year}{2020}\natexlab{}.
\newblock \showarticletitle{{DHPA:} Dynamic Human Preference Analytics Framework: {A} Case Study on Taxi Drivers' Learning Curve Analysis}.
\newblock \bibinfo{journal}{\emph{{ACM} Trans. Intell. Syst. Technol.}} \bibinfo{volume}{11}, \bibinfo{number}{1} (\bibinfo{year}{2020}), \bibinfo{pages}{8:1--8:19}.
\newblock


\bibitem[Park et~al\mbox{.}(2020)]%
        {whatdriverwants}
\bibfield{author}{\bibinfo{person}{So~Yeon Park}, \bibinfo{person}{Dylan~James Moore}, {and} \bibinfo{person}{David Sirkin}.} \bibinfo{year}{2020}\natexlab{}.
\newblock \showarticletitle{What a Driver Wants: User Preferences in Semi-Autonomous Vehicle Decision-Making}. In \bibinfo{booktitle}{\emph{{CHI} '20: {CHI} Conference on Human Factors in Computing Systems, Honolulu, HI, USA, April 25-30, 2020}}, \bibfield{editor}{\bibinfo{person}{Regina Bernhaupt}, \bibinfo{person}{Florian~'Floyd' Mueller}, \bibinfo{person}{David Verweij}, \bibinfo{person}{Josh Andres}, \bibinfo{person}{Joanna McGrenere}, \bibinfo{person}{Andy Cockburn}, \bibinfo{person}{Ignacio Avellino}, \bibinfo{person}{Alix Goguey}, \bibinfo{person}{Pernille Bj{\o}n}, \bibinfo{person}{Shengdong Zhao}, \bibinfo{person}{Briane~Paul Samson}, {and} \bibinfo{person}{Rafal Kocielnik}} (Eds.). \bibinfo{publisher}{{ACM}}, \bibinfo{pages}{1--13}.
\newblock


\bibitem[Schneider et~al\mbox{.}(2021)]%
        {DBLP:conf/chi/SchneiderHRGTG21}
\bibfield{author}{\bibinfo{person}{Tobias Schneider}, \bibinfo{person}{Joana Hois}, \bibinfo{person}{Alischa Rosenstein}, \bibinfo{person}{Sabiha Ghellal}, \bibinfo{person}{Dimitra Theofanou{-}F{\"{u}}lbier}, {and} \bibinfo{person}{Ansgar R.~S. Gerlicher}.} \bibinfo{year}{2021}\natexlab{}.
\newblock \showarticletitle{ExplAIn Yourself! Transparency for Positive {UX} in Autonomous Driving}. In \bibinfo{booktitle}{\emph{{CHI} '21: {CHI} Conference on Human Factors in Computing Systems, Virtual Event / Yokohama, Japan, May 8-13, 2021}}, \bibfield{editor}{\bibinfo{person}{Yoshifumi Kitamura}, \bibinfo{person}{Aaron Quigley}, \bibinfo{person}{Katherine Isbister}, \bibinfo{person}{Takeo Igarashi}, \bibinfo{person}{Pernille Bj{\o}rn}, {and} \bibinfo{person}{Steven~Mark Drucker}} (Eds.). \bibinfo{publisher}{{ACM}}, \bibinfo{pages}{161:1--161:12}.
\newblock


\bibitem[Song et~al\mbox{.}(2013)]%
        {DBLP:conf/models/SongBCC13}
\bibfield{author}{\bibinfo{person}{Hui Song}, \bibinfo{person}{Stephen Barrett}, \bibinfo{person}{Aidan Clarke}, {and} \bibinfo{person}{Siobh{\'{a}}n Clarke}.} \bibinfo{year}{2013}\natexlab{}.
\newblock \showarticletitle{Self-adaptation with End-User Preferences: Using Run-Time Models and Constraint Solving}. In \bibinfo{booktitle}{\emph{Model-Driven Engineering Languages and Systems - 16th International Conference, {MODELS} 2013, Miami, FL, USA, September 29 - October 4, 2013. Proceedings}} \emph{(\bibinfo{series}{Lecture Notes in Computer Science}, Vol.~\bibinfo{volume}{8107})}, \bibfield{editor}{\bibinfo{person}{Ana Moreira}, \bibinfo{person}{Bernhard Sch{\"{a}}tz}, \bibinfo{person}{Jeff Gray}, \bibinfo{person}{Antonio Vallecillo}, {and} \bibinfo{person}{Peter~J. Clarke}} (Eds.). \bibinfo{publisher}{Springer}, \bibinfo{pages}{555--571}.
\newblock


\bibitem[Soules and Ganger(2003)]%
        {DBLP:conf/hotos/SoulesG03}
\bibfield{author}{\bibinfo{person}{Craig A.~N. Soules} {and} \bibinfo{person}{Gregory~R. Ganger}.} \bibinfo{year}{2003}\natexlab{}.
\newblock \showarticletitle{Why Can't {I} Find My Files? New Methods for Automating Attribute Assignment}. In \bibinfo{booktitle}{\emph{Proceedings of HotOS'03: 9th Workshop on Hot Topics in Operating Systems, May 18-21, 2003, Lihue (Kauai), Hawaii, {USA}}}, \bibfield{editor}{\bibinfo{person}{Michael~B. Jones}} (Ed.). \bibinfo{publisher}{{USENIX}}, \bibinfo{pages}{115--120}.
\newblock


\bibitem[Tener and Lanir(2022)]%
        {DBLP:conf/chi/TenerL22}
\bibfield{author}{\bibinfo{person}{Felix Tener} {and} \bibinfo{person}{Joel Lanir}.} \bibinfo{year}{2022}\natexlab{}.
\newblock \showarticletitle{Driving from a Distance: Challenges and Guidelines for Autonomous Vehicle Teleoperation Interfaces}. In \bibinfo{booktitle}{\emph{{CHI} '22: {CHI} Conference on Human Factors in Computing Systems, New Orleans, LA, USA, 29 April 2022 - 5 May 2022}}, \bibfield{editor}{\bibinfo{person}{Simone D.~J. Barbosa}, \bibinfo{person}{Cliff Lampe}, \bibinfo{person}{Caroline Appert}, \bibinfo{person}{David~A. Shamma}, \bibinfo{person}{Steven~Mark Drucker}, \bibinfo{person}{Julie~R. Williamson}, {and} \bibinfo{person}{Koji Yatani}} (Eds.). \bibinfo{publisher}{{ACM}}, \bibinfo{pages}{250:1--250:13}.
\newblock


\bibitem[Wohlrab et~al\mbox{.}(2022)]%
        {DBLP:conf/seams/WohlrabMV22}
\bibfield{author}{\bibinfo{person}{Rebekka Wohlrab}, \bibinfo{person}{R{\^{o}}mulo Meira{-}G{\'{o}}es}, {and} \bibinfo{person}{Michael Vierhauser}.} \bibinfo{year}{2022}\natexlab{}.
\newblock \showarticletitle{Run-Time Adaptation of Quality Attributes for Automated Planning}. In \bibinfo{booktitle}{\emph{International Symposium on Software Engineering for Adaptive and Self-Managing Systems, {SEAMS} 2022, Pittsburgh, PA, USA, May 22-24, 2022}}, \bibfield{editor}{\bibinfo{person}{Bradley~R. Schmerl}, \bibinfo{person}{Martina Maggio}, {and} \bibinfo{person}{Javier C{\'{a}}mara}} (Eds.). \bibinfo{publisher}{{ACM/IEEE}}, \bibinfo{pages}{98--105}.
\newblock


\bibitem[Yang and Mesbah(2013)]%
        {routechoice}
\bibfield{author}{\bibinfo{person}{C. Yang} {and} \bibinfo{person}{M. Mesbah}.} \bibinfo{year}{2013}\natexlab{}.
\newblock \showarticletitle{Route choice behaviour of cyclists by stated preference and revealed preference.}
\newblock \bibinfo{journal}{\emph{Australasian Transport Research Forum, ATRF 2013 - Proceedings.}} (\bibinfo{year}{2013}).
\newblock


\end{thebibliography}

\appendix
\section{Appendix}
\label{sec:appendix}
\subsection{Complaint Encoding into Fitness Function}
\label{sec:ga}

We leverage the genetic algorithms (GA) to facilitate the preference update process. In this GA-based approach: 1)each preference configuration is treated as an individual within the GA (two representative individuals can be described as $p_1=\langle w_1,w_2,...,w_n \rangle$ and $p_2=\langle w_1',w_2',...,w_n' \rangle$); 2) every weight, denoted by $w_i$, acts as a gene within the said individual; 3) A group of these individuals comes together to form what is termed a "population". By adopting this structure, the GA allows for a systematic and evolutionary approach to updating and optimizing preference vectors based on a given set of criteria.


Within GA, the fitness of each individual in the population is assessed in every iteration.The primary objective of updating preferences in our study is to augment user satisfaction and minimize user complaints, leading us to embed user complaints into the fitness function.


The function is formulated as: $f(p)=\lambda_1f_1+\lambda_2f_2+\lambda_3f_3$, where $\lambda_1$, $\lambda_2$, $\lambda_3$ are positive coefficient parameters, and $p=\langle w_1,...,w_n\rangle$ is the vector of preferences. 
The functions $f_1$, $f_2$, and $f_3$ quantitatively characterize the evaluation of three distinct aspects:
(1) \emph{Preference Divergence.} Minimizing the discrepancy in preferences pre and post-update while maintaining user satisfaction. The basic idea behind this discrepancy is that each individual changes his preference in a small step within a certain time frame~\cite{driverPreference}. 
(2) \emph{Complaint Avoidance.} Assessing if the optimal path, derived from the given preference, traverses any complained-about states - it ideally shouldn't.
(3) \emph{Implicit Constraints.} Addressing constraints tied to the following four complaint categories (General Discontent, Specific Discontent, General Preference Discontent), like projecting an attribute weight increase or one attribute's weight overtaking another's. 
The meaning of these four implicit complaints is as follows: 1. \emph{General Discontent}: E.g., "I don't like this road." 2. \emph{Specific Discontent}: E.g., "the road is too bumpy" or "it's too noisy." 3. \emph{General Preferences Discontent}: E.g., "road condition takes precedence over trip efficiency." 4. \emph{Specific Preferences Discontent}: E.g., "The weight for road condition should be 0.8!".
It's evident that both "Preference Divergence" and "Complaint Avoidance" are universally relevant across complaint categories. As such, both $f_1$ and $f_2$ remain consistent throughout.  However, given the inherent uniqueness of each complaint type, designing specific $f_3$ functions for each becomes essential.

Function $f_1$ represents the cosine similarity between two preference configurations, $p$ and $p'$.  $f_2$ accounts for user complaint avoidance, with $\rho_1$ acting as a negative penalty parameter, $\tau$ being determined by $\text{PATH\_FINDING}(map, p)$, and $\mathbb{C}$ symbolizes a set of states.

\begin{minipage}{.67\linewidth}
\begin{equation}
\label{eq:sim}
f_1=-\textsc{CosSim}(p,p')\ = -\quad\dfrac{p \wedge p'}{||p||\times ||p'||} 
= -\quad\dfrac{\quad\sum\limits_{i=1}^{n} w_i\times w_i'}{\sqrt{\sum\limits_{i=1}^{n} w_i^2}\times \sqrt{\sum\limits_{i=1}^{n} w_i'^2}}
\end{equation}
  \end{minipage}
\begin{minipage}{.3\linewidth}
    \begin{equation}
    \label{eq:finess2}
\begin{aligned}
    f_2
    =\begin{cases}
    \rho_1,  \text{\ \ \ if $\mathbb{C}\cap \tau \neq \emptyset$}\\
    0, \text{\ \ \ \ \ otherwise}
    \end{cases}
    \end{aligned}
\end{equation}
\end{minipage}%



For the design of $f_3$ based on complaint categories, various considerations emerge. For General Discontent complaints, like "dislike of the road", $f_3$ is straightforwardly set to a constant value of 0, given its absence of supplemental constraints. Diving into Specific Discontent, we categorize the complaints by their attribute order: $\langle$ Road Condition, Efficiency, Aesthetic Appeal $\rangle$. For instance, complaints like "excessive noise!" maps to attribute $id = 3$, "excessive bumpiness!" to $id = 1$,  and "excessive distance" to $id = 2$. The structure of $f_3$ in this category leverages a negative penalty parameter, $\rho_2 < 0$, coupled with a fuzzy logic representation. This induces a transition between conditions $w_{id}\leq w'_{id}$ and $w_{id}>w'_{id}$. Additionally, the bias  $b$ is derived from $b=\min(\phi,1-w'_{id})$ where $\phi$, a hyperparameter, represents  the size of this transition, usually lying between 0.1 and 0.2. A detailed representation is found in Eq.(\ref{eq:fitness3}).

\begin{equation}
\label{eq:fitness3}
\begin{aligned}
    f_3
    =\begin{cases}
    \rho_2,  \text{\quad \quad \quad \quad \quad \quad \quad  if $w_{id}\in [0,w'_{id}]$}\\
    \rho_2\big(\frac{w'_{id}-w_{id}}{b}+1\big), \text{\quad \  if $w_{id}\in (w'_{id},w'_{id}+b]$}\\
    0, \text{\quad \quad \quad \quad \quad \quad \quad \ if $w_{id}\in (w'_{id}+b,1]$}
    \end{cases}
    \end{aligned}
\end{equation}


When addressing General Preference Discontent, two pivotal attributes, $id1$ and $id2$, emerge from the complaint, with the former overshadowing the latter in significance. This is captured succinctly in Eq.(\ref{eq:fitness4}). Lastly, for Specific Preference Discontent, the complaint pinpoints a particular attribute, $id$, and stipulates an ideal preference value, $w_{opt}$. The expression for $f_3$ under this complaint type is as shown in Eq.(\ref{eq:fitness5}).

\begin{minipage}{.45\linewidth}
\begin{equation}
\label{eq:fitness4}
\begin{aligned}
    f_3
    =\begin{cases}
    \rho_2,  \text{\quad \quad  if $w_{id1}<w_{id2}$}\\
    0, \text{\ \ \quad \quad  if $w_{id1}\geq w_{id2}$}
    \end{cases}
    \end{aligned}
\end{equation}
  \end{minipage}
\begin{minipage}{.45\linewidth}
 \begin{equation}
\label{eq:fitness5}
\begin{aligned}
    f_3
    =\begin{cases}
    \rho_2,  \text{\quad \quad if $w_{id}=w_{opt}$}\\
    0, \text{\quad \quad \ \  if $w_{id}\neq w_{opt}$}
    \end{cases}
    \end{aligned}
\end{equation}
\end{minipage}%

The condition $\rho_1\ll \rho_2 < 0$ is essential 
because complaint avoidance offers a critical user feedback, indicating the system to steer clear of complaint points. This feedback must always be prioritized, while other considerations from implicit constraints and preference divergence can be treated with more discretion. 

\subsection{GA-Based Preference Adaptation}

The GA-based preference adaptation starts with a random population initialization and then progresses through successive generations using selection, crossover, and mutation operations. The process continues until a termination criterion is met, either by exceeding a set number of iterations or when the population's best fitness remains unchanged for a certain period.

\emph{Selection operation}:
After calculating the fitness of each individual in the population, individuals are selected (with potential repetition) based on the following probability: 
\begin{equation}
    P(p)=\frac{f(p)-p_{min}}{\sum_{p'\in Pop}\big(
    f(p)-p_{min}
    \big)}
\end{equation}
where $p_{min}=\min_{p'\in Pop}f(p')$, and $Pop$ is the population, and the operation could be implemented with Roulette-wheel selection.

\begin{figure}
    \centering
    \includegraphics[width=1.01\linewidth]{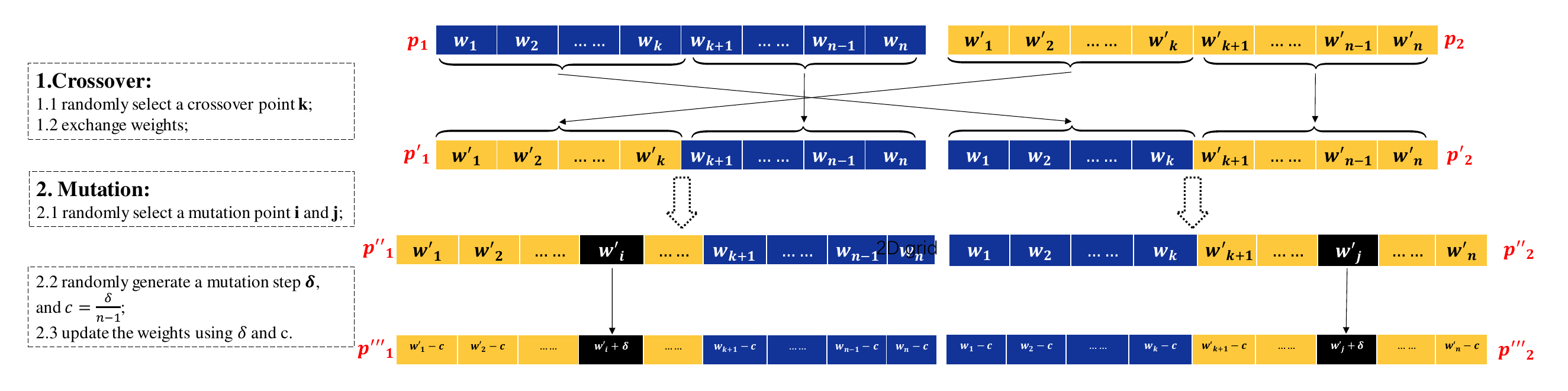}
    \caption{An Illustration of Crossover and Mutation Operations in GA-based Preference Update.}
    \label{fig:ga-alg}
\end{figure}

\emph{Crossover operation}:
The selected individuals constitute a new population. 
Within this new population, individuals are paired two by two for crossover operations. 
The operation process is depicted in Fig.\ref{fig:ga-alg}: given two individuals $p_1$ and $p_2$, an initial crossover point $k\in \{1,2,...,n\}$ is selected randomly. 
Then, the genes from 1 to $k$ of individual $p_2$ (i.e., $w'_1$ to $w'_k$) and the genes from $k+1$ to $n$ of individual $p_1$ (i.e.,$w_{k+1}$ to $w_n$) are combined to form a new individual represented as $p'_{1}$. 
Likewise, the genes from 1 to $k$ of $p_1$ (i.e., $w_1$ to $w_k$) and the genes from $k+1$ to $n$ of $p_2$ (i.e., $w'_{k+1}$ to $w'_n$) are combined to form a new individual represented as $p_{2}'$.
To ensure the legitimacy of $p'_1$ and $p'_2$, if a crossover results in the condition $\sum_{i=1}^k{w_i'} + \sum_{i=k+1}^n{w_i}>1$, then $p_1'$ is directly set to $p_1$ and $p_2'$ is directly set to $p_2$.

\emph{Mutation operation}:
Upon obtaining the new individuals after the crossover, a mutation operation is then performed. The procedure for the operation is illustrated in Fig. \ref{fig:ga-alg}:
for each individual, a mutation point $i\in\{1,2,...,n\}$ and a mutation step $\delta$ are selected randomly.
Subsequently, the individual's gene \( w_i \) is modified to \( w_i + \delta \), while the other genes are modified to \( w-\frac{\delta}{n-1} \). 
Let $w_k$ denote the weights before mutation, and $w'_k$ represent the weights after mutation.
If the mutation results in $\exists k\in \{1,2,...,n\}.w'_k <0$ or $\exists k\in \{1,2,...,n\}. w'_k >1$, then $\delta$ needs to be trimmed as follows: $\delta=\min(\delta,\min_{k\in\{1,2,...,n\}}|w_k-0|,\min_{i\in\{1,2,...,n\}}|w_k-1|)$.

\end{document}